\documentclass[authoryear,12pt]{elsarticle}

\usepackage{amssymb}
\usepackage{amsfonts}
\usepackage{amsmath}
\usepackage{mathrsfs}
\usepackage{latexsym}
\usepackage{natbib}
\usepackage{epsfig}

\usepackage{lineno}

\usepackage[utf8]{inputenc}
\usepackage{lmodern,textcomp}

\usepackage{fancyhdr}
\journal{a journal}

\begin{document}

\begin{frontmatter}

\title{Optimal forest rotation under carbon pricing and forest damage risk}

\author[Aalto]{Tommi Ekholm\corref{cor1}}
\address[Aalto]{Finnish Meteorological Institute, Erik Palmenin aukio 1, P.O. BOX 503, 00101 Helsinki, Finland}
\cortext[cor1]{Corresponding author. Tel: +358 40 775 4079}
\ead{tommi.ekholm@fmi.fi}


\begin{abstract}
Forests will have two notable economic roles in the future: providing renewable raw material and storing carbon to mitigate climate change. The pricing of forest carbon leads to longer rotation times and consequently larger carbon stocks, but also exposes landowners to a greater risk of forest damage. This paper investigates optimal forest rotation under carbon pricing and forest damage risk. I provide the optimality conditions for this problem and illustrate the setting with numerical calculations representing boreal forests under a range of carbon prices and damage probabilities.  The relation between damage probability and carbon price towards the optimal rotation length is nearly linear, with carbon pricing having far greater impact. As such, increasing forest carbon stocks by lengthening rotations is an economically attractive method for climate change mitigation, despite the forest damage risk. Carbon pricing also increases land expectation value and reduces the economic risks of the landowner. The production possibility frontier under optimal rotation suggests that significantly larger forests carbon stocks are achievable, but imply lower harvests. However, forests’ societally optimal role between these two activities is not yet clear-cut; but rests on the future development of relative prices between timber, carbon and other commodities dependent on land-use.
\end{abstract}

\begin{keyword}
Optimal forest rotation \sep climate change \sep carbon pricing \sep forest damage
\end{keyword}

\end{frontmatter}


\section{Introduction}

Meeting ambitious climatic goals -- such as pursuing efforts to limit the temperature increase to 1.5$^\circ$C, as prescribed in the Paris Agreement -- is likely to require removing significant amounts of carbon dioxide from the atmosphere \citep{Rogelj2018}.
This heightened challenge in climate change mitigation has prompted stronger interest in enhancing forest carbon sinks \citep[see][for a review]{Sedjo2012}. 
Due to the vast size of the forest ecosystems, large potential exists in both capturing and storing atmospheric carbon.
Forests could, therefore, provide a significant contribution to the grand effort needed to mitigate climate change and meet the Paris Agreement targets \citep[see e.g.][]{Griscom2017, Favero2017}.

While an extensive expansion of forest carbon stocks might be possible from the perspective of natural sciences, it is not evident that such would be economically optimal.
In an ideal economic setting, all flows of carbon to and from forests should be priced with the social cost of carbon.
A policy that imposes a price for forest carbon -- e.g. through subsidies and taxes for the landowner -- would provide an efficient incentive to expand the forest carbon stock. This would generally lead to longer rotation times in managed forests \citep{vanKooten1995}. Further, the effect is greatly amplified if the price of carbon increases over time, which is a common result from long-term economic assessments on optimal mitigation strategies \citep{Ekholm2016}.

By lengthening the rotation, however, the landowner is exposed to a higher risk that a fire, storm or other damage destroys the stand before the trees are harvested. 
It has been shown that the presence of a forest damage risk shortens the optimal rotation \citep{Reed1984}, although this analysis did not account for forest carbon pricing. 
By shortening the optimal rotation, the forest damage risk leads to lower forest carbon stocks and a lower potential for mitigating climate change. Although \citet{Amacher2005} noted that under a possibility for fire management, a higher fire risk could imply longer rotations; this was accompanied by lower planting densities, thus potentially also implying lower forest carbon stocks. 

Additional to the loss of harvestable wood, forest damages are detrimental towards the climatic aims. Damages disrupt forest growth, and therefore the trees' capability to sequester atmospheric carbon; and release some of the stored carbon back to the air, particularly in the case of forest fires.
These factors can attenuate the impact of a carbon pricing policy on forest management.
A further motivation for considering the damage risk is that climate change itself is likely to increase the risk level in the future.

For understanding how large a contribution forests might have towards mitigating climate change, were proper incentives put in place, it is of importance to analyze optimal rotation in the presence of carbon pricing and forest damage risk \citep{Stainback2004,Stollery2005,Daigneault2010,Couture2011}.
This will answer which of the counteracting forces -- carbon pricing lengthening and damage risk shortening the optimal rotation -- dominates optimal forest management.
The optimal rotation approach can illustrate how largely the damage risk might obviate the economic potential for forests to drain carbon, and also how the relative pricing between timber and carbon affects the supply of wood and mitigation.
Further, one can also investigate how carbon pricing and the damage risk affect the returns and economic risks for the landowner.

This paper presents the first-order optimality conditions for forest rotation length under carbon pricing and forest damage risk. 
In essence, I combine the approaches of \citet{Reed1984} for forest damage risk and \citet{vanKooten1995} for carbon pricing. 
Using this model, I present the optimal rotation age, land expectation value, the risk in returns, and the production possibility frontier between carbon stocks and timber yield \citep{Pingoud2018}.

A number of authors have analyzed the relationship between carbon pricing and damage risk using numerical optimization methods.
\citet{Stainback2004} optimize the land expectation value directly in a relatively similar problem setting, while \citet{Couture2011} and \citet{Daigneault2010} used stochastic dynamic programming, taking respectively also thinnings and risk preferences into account.

The model I employ in this paper is closest to that of \citet{Stainback2004}, but more general in terms of the stochastic process underlying forest damages, the pricing of timber harvested at different ages, and the carbon dynamics after a harvest or a damage event.
Moreover, the recursive formulation used here allows presenting the model in a more compact form. 
The numerical examples portraying boreal forests involve more realistic representation of carbon retention in dead wood and products, with numerical estimates based on empirical literature.
I also present results on unexplored aspects, such as the economic risk and the production possibility frontier between wood production and carbon storage.

\section{Determining the optimal rotation length}

The problem setting aims to find the rotation length $T$ that maximizes the expected net revenues from an infinite chain of rotations when starting from bare land. Changes in carbon stock are priced with a constant price $P_c$ (in €/t$_{CO2}$), analogously to the approach of \citet{vanKooten1995}. The price of harvested timber from final fellings is $P_f(T)$ (in €/m$^3$), which depends on what age $T$ the stand is harvested. 
%
The timber value $P_f(T)$ does not, however, cover the value of carbon stored in the harvested wood. Instead, the gradual release of carbon from the wood product pool is priced separately from the timber price. Similar approach is used for the decay of harvest residues and dead organic matter after forest damage.
Future income and costs are discounted with rate $r$.

At each point of time, the considered forest area is subject to a risk of forest damage. In case the forest damage occurs, some of the carbon stock can be released to the atmosphere, while some part of the wood can be salvaged and sold. New rotation begins after final fellings or damage, and incurs a regeneration cost $R$.

Let $Z_1, Z_2, ...$ be a sequence of random variables that represent the stand ages at which a damage event would occur for each subsequent rotation. If final fellings occur at age $T$, the $n^{th}$ rotation ends in harvest if $T \leq Z_n$. Otherwise the damage destroys the stand and a new rotation follows.\footnote{\citet{Xu2016} have provided an extension that allows multiple disturbances in one rotation, whereby the stand is not lost completely in a damage event.} Assume that the random variables are independent and identically distributed. Denote the probability density for $Z_n = t$ with $p(t)$. This approach is very similar to that of \citet{Reed1984}, but here the function that defines the probability is generic, allowing the probability to depend on the stand age. It is worth to note, however, that the stand age is used merely as a proxy for the evolution of the stand’s physical characteristics, like density.

To determine the land expectation value in this setting, let us first split the net revenue calculation from a single rotation into two mutually exclusive cases. $D(t)$ indicates net revenues if forest damage occurs at time $t$, with $t < T$:
\begin{equation}
\label{eq:D}
D(t) = \alpha  P_c \int_0^t e^{-r \tau } v'(\tau ) \, d\tau - e^{-r t} \left((1-\gamma) \alpha P_c v(t) + R \right)
\end{equation}
$H(T)$ indicates net revenue if no damage occurs, and final fellings take place at time $T$:
\begin{equation}
\label{eq:H}
H(T) = \alpha P_c \int_0^T e^{-r \tau } v'(\tau ) \, d\tau + e^{-r T} \left( \left(P_F(T)- (1-\beta) \alpha P_c\right) v(T) - R\right)
\end{equation}
In these equations, $v(t)$ denotes the stem volume at age $t$ (in m$^3$/ha)\footnote{All values are here given in SI or compatible units. t$_C$ stands for tonnes of carbon and t$_{CO_2}$ for tonne of carbon dioxide, which can be converted to each other using the ratio of molecular weights.} and $\alpha$ is the total carbon content of living biomass per stem volume (in t$_{CO_2}$/m$^3$, as $P_c$ is given in terms of €/t$_{CO2}$), $\gamma$ and $\beta$ are respectively the fractions of carbon remaining stored in cases of forest damage and final fellings. That is, the forest owner is required to pay for the fraction $(1-\gamma)$ or $(1-\beta)$ of the trees' carbon content following forest damage or harvest.

The expressions $D(t)$ and $H(T)$ cover only a single rotation. The objective is, however, to maximize the expected net present value from an infinite chain of rotations. As the problem setting is identical for each rotation -- prices and other parameters are assumed to remain constant over time --  it is possible to use a simple recursive formulation.

Let $V(T)$ be a value function, portraying the expected net present value subject to the rotation length $T$.
Then, the expected land value for rotation length $T$ can be expressed as
\begin{equation}
\label{eq:recursive}
V(T) = \int_0^T p(t) \left(D(t) + e^{-r t}V(T) \right) \, dt+\int_T^{\infty } p(t) \left(H(T) + e^{-r T}V(T)\right) \, dt
\end{equation}
This formulation incorporates the expected value of subsequent rotations recursively through the expected land value $V(T)$, discounted to the start of the first rotation. 
Upon rearranging, $V(T)$ can be presented as a function of the optimal rotation length:
\begin{equation}
\label{eq:valuefunction}
    V(T)=\frac{\int_0^{T} D(t) p(t) \, dt+H(T) \int_{T}^{\infty } p(t) \, dt}{1 - \int_0^{T} p(t) e^{-r t} \, dt - e^{-r T} \int_{T}^{\infty } p(t) \, dt}
\end{equation}

This formula resembles the expression for bare land value in a non-probabilistic setting -- i.e. without the possibility of forest damage -- only that the net revenues in the dividend and the recursion factor in the divisor are now calculated in the form of expected values.
The first-order optimality conditions in this general setting could be stated by finding the rotation age $T^*$ for which the value function  \eqref{eq:valuefunction} has a derivative of zero. 

So far the damage probability $p(t)$ has been described to be an arbitrary function conditional on the stand age. 
To present a more concrete case, the annual damage probability is assumed in the following to remain constant over the stand's lifetime. Consequently, the damages follow a Poisson process and occur at intervals that are exponentially distributed:
\begin{equation}
\label{eq:probability}
    p(t) = \lambda  e^{-\lambda  t}
\end{equation}
where $\lambda$ denotes the average rate of forest damages per year.

Under this probability assumption, upon some algebraic manipulation on the first-order condition $V'(T^*) = 0$, one can arrive at the following expression
\begin{eqnarray}
\label{eq:FOC}
&&
\Big( \left(1-e^{-(\lambda+r)T^*}\right) P_F'(T^*) - (\lambda+r) P_F(T^*)
\\ && \nonumber
\hspace{7mm} -\alpha P_c \left(r(\beta-1) + \lambda(\beta-\gamma) - \lambda(1-\gamma) e^{-(\lambda +r)T^*}\right)  \Big) v(T^*)
\\ && \nonumber
+ \left(1-e^{-(\lambda +r) T^* }\right) \left(\alpha  \beta  P_c+P_F(T^*)\right) v'(T^*) 
\\ && \nonumber
-\alpha P_c (\lambda +r) e^{-\lambda  T^*} \int_0^{T^*} e^{-r \tau } v'(\tau ) \, d\tau 
\\ && \nonumber
-(\lambda +r) \int_0^{T^*} \lambda  e^{-\lambda t} \left(  \alpha P_c \int_0^t e^{-r \tau } v'(\tau ) \, d\tau - e^{-r t} \left(\alpha P_c (1-\gamma) v(t)+R\right)  \right) dt
\\ && \nonumber
+ \left(r + \lambda e^{-(\lambda +r)T^*}\right)R = 0,
\end{eqnarray}
which allows to solve the optimal rotation length $T^*$ numerically.

As the problem setting is a generalization from the problem settings of both \citet{Reed1984} and \citet{vanKooten1995}, equation \eqref{eq:FOC} simplifies to the optimality conditions of these cases with appropriate parameters. Setting $P_c = 0$ eliminates carbon pricing, and with a constant timber price $P_F(T) = P_F$ the equation \eqref{eq:FOC} simplifies to the first-order conditions proposed by \citet{Reed1984}. Similarly, setting $\lambda = 0$, $R = 0$ and $P_F(T) = P_F$ eliminates the damage risk and regeneration costs, making \eqref{eq:FOC} correspond the formula presented by \citet{vanKooten1995}. Substitutions $P_c = 0$, $P_F(T) = P_F$ and $\lambda = 0$ together simplify \eqref{eq:FOC} to the first-order condition of the Faustmann problem.

%
%
Last, one should note that the parameters $\gamma$ and $\beta$ provide considerable flexibility in determining the carbon costs due to damage or harvest, respectively. 
In the numerical examples of Section \ref{sec:numerical}, I assume that carbon is released gradually to the atmosphere after harvest or a damage event, and that this gradual decay is priced with $P_c$. The net present value of this can be represented as a cost to the forest owner during the harvest or damage through $\beta$ and $\gamma$. Although in the case of harvests, these costs would be borne by some other actor in the economy, efficient markets would pass this cost to the price of timber, and this affects the forest-owners decision-making. 
As $P_f(t)$ was assumed to exclude the value of carbon embodied in the sold timber, this lower price due to the carbon cost can be captured through $\beta$.

Moreover, $\gamma$ can be used to represent that a fraction $\delta$ of the timber value can be salvaged and sold after a damage event, including the salvage cost and lower value of the salvaged wood. Assume that a fraction $\tilde{\gamma}$ of the carbon is retained after damage. Then, one can write $\gamma = \tilde{\gamma}+\frac{\delta P_F}{\alpha P_c}$. By inserting this into \eqref{eq:D}, one arrives at two terms representing the carbon costs and salvage value  after a damage event.
One can also set $\gamma = 0$ if no payments are required for the carbon release from forest damage.

\section{Numerical cases}
\label{sec:numerical}

To illustrate, I calculate the optimal rotation for Scots pine and Norway spruce stands in Southern Finland under a range of carbon prices and damage risks. The calculated cases depart from reality in that the problem formulation omits thinnings, although they are a common practice in Finnish forest management;but nevertheless provide insights into the effects of carbon pricing and damage risks on optimal rotation in boreal forests. The interested reader can refer to e.g. \citet{Pohjola2007} for an analysis with carbon pricing and thinnings. 

\subsection{Setup}
\label{sec:setup}

\begin{figure}[ht!]
  \centering
    \includegraphics[width=\textwidth]{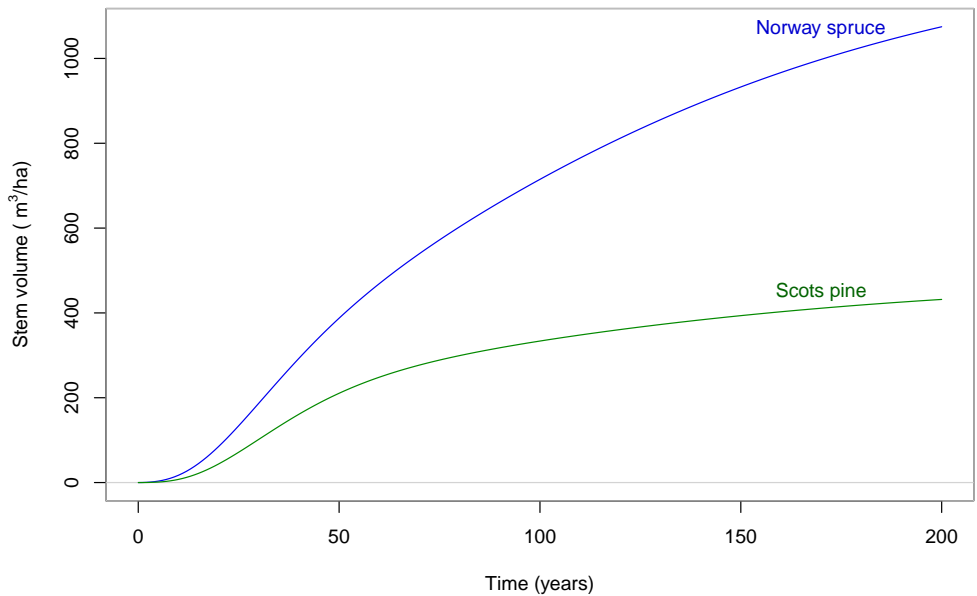}
      \caption{Stem volume as a function of stand age $v(t)$ for illustrative Scots pine and Norway spruce stands in Southern Finland.}
      \label{StemVolume}
\end{figure}

The stem volume is presented in Figure \ref{StemVolume} as a function of stand age. The functional representation is determined through $v'(t) = v_1 t e^{v_2 t} + v_3 t^3 e^{v_4 t}$, which allows flexible fit and a closed-form solution for both $v(t)$ and when integrating $e^{-r t} v'(t)$, as is done in \eqref{eq:FOC}. The parameter values are given in Table \ref{parameters}. The conversion from stem volume to carbon content ($\alpha$), is from \citet{Lehtonen2004}, assuming a 50\% carbon content of biomass. The discount rate is fixed at 3\% and regeneration cost at zero in all calculations. 

The price of timber depends on the stand age, reflecting that younger stands yield more low-price pulpwood and older stands heavier logs that are priced hgher. The price is assumed to increase from zero to $P_{F,max}$=60~€/m$^3$, according to $P_f(t) = (\mu t)^2 e^{\mu t}/(1+(\mu t)^2 e^{\mu t}) P_{F,max}$, where $\mu=0.015$. 
This parametrisation is based loosely on energy wood, pulpwood and sawlog prices in Southern Finland during the last 10 years, and used for illustrative purposes.
It should be noted that optimal rotation lengths for different timber prices can be obtained from the results calculated here by scaling $P_f(t)$ and $P_c$ proportionally.

I consider here two types of forest damage: storm and fire. 
Carbon is released with differing dynamics during and after a storm or forest fire, for which I provide estimates based on earlier empirical research. Analogous estimates are made for carbon retention if the trees are felled for long-lived wood products. The approach could be parametrized for other damages, e.g. insect infestations, in subsequent research.

The shares of carbon released after damage ($1-\gamma$) or harvest ($1-\beta$) are calculated as the discounted amount of carbon lost each year after the damage event or harvest. When multiplied with the carbon price $P_c$ -- as is done in equations \eqref{eq:D} and \eqref{eq:H} -- these factors portray the present value of payments from the gradual loss of the carbon stock.

The shares of carbon stored in different tree compartments -- roots, stem, branches and foliage -- are from \citet{Lehtonen2004} and their decay rates from \citet{Repo2011}. The soil carbon stock due to natural litter production and decay is not considered here. After a storm, the whole tree biomass is assumed to be left on site. The stem and stump are assumed to decay with the rate of 20 cm stumps in \citet{Repo2011}, the branches with 2 cm diameter rate, and the foliage rapidly. For fire, based on \citet{Liski1998}, I assume that 100\% of needles, 75\% of branches and 25\% of stemwood is burnt; with the remaining fraction decaying as above.

For harvested wood products, I assume the wood is used in sawmills and ends up in construction, using numbers from \citet{Pingoud2001}. From the stemwood carbon stock, 56\% is released at the sawmill; and of the remaining fraction, 50\% ends up in long timeframe and 15\% in medium timeframe products, the rest being released. Harvest residues are assumed to remain on site and to decay as above.

\begin{figure}[htb!]
  \centering
    \includegraphics[width=\textwidth]{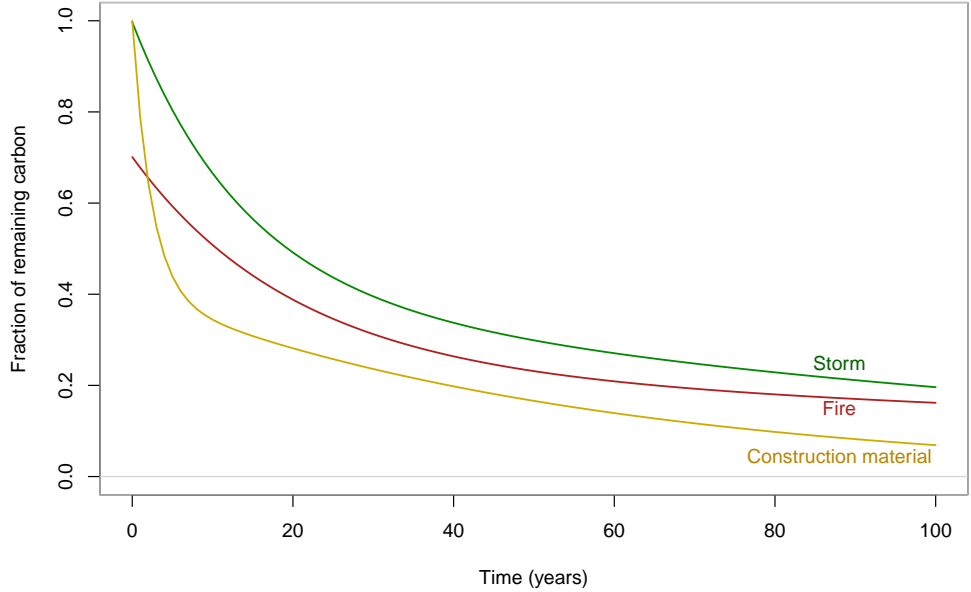}
      \caption{The dynamics of carbon stock remaining after a storm, fire or harvest for sawmill products.}
      \label{dynamics}
\end{figure}

The resulting carbon stock dynamics are presented in Figure \ref{dynamics}. The values of $\gamma$ and $\beta$ -- when discounted with a 3\% rate – are presented in Table \ref{parameters}.

Two parameters for which it is harder to provide reliable estimates are the carbon price and damage probability. A recent sensitivity analysis regarding factors underpinning the social cost of carbon found a range from 10~\$/t$_{CO_2}$ to 300~\$/t$_{CO_2}$ defensible \citep{Ekholm2018}. With a discount rate of 3\%, which is same as used here, the carbon price range extends to slightly over 100~\$/t$_{CO_2}$. 

While statistics on the historical frequency of forest damages do exist, they do not reflect the future risk level in a warmer climate.
To have some point of reference, nevertheless, one could look at the exceptionally vast wildfires in Sweden during summer 2018. A result of a heatwave throughout the Northern Hemisphere, reported 25 000 hectares of forests were affected. While extensive, the affected area corresponds to only 0.1\% of Sweden's productive forest area. 
\citet{Lehtonen2016} estimated that the number of fires in Finland could almost double within the century under moderate climate change.
The risk is yet highly dependent on local conditions, and far higher damage probabilities have been presented for temperate forests \citep{Stainback2004}. Climate change is, nevertheless, projected to increase particularly the fire risks throughout the planet \citep{Seidl2017}.

Based on these considerations, I calculate the results for a range of carbon prices and annual probabilities of damage, from 0 to 100~€/t$_{CO_2}$ and 0\% to 1\% respectively.


\begin{table}[ht!]
\linespread{2.2}
\caption{Species-specific parameter values for the illustrative calculations.}
    \hspace{-15mm}
    \begin{center}
    \begin{tabular}{ll|cc}    
        && Scots pine & Norway spruce \\
        \hline
        Carbon content per stem volume & $\alpha$ (t$_{CO_2}$/m$^3$) & 1.29 & 1.36 \\
        Carbon stored after fire & $\gamma_{fire}$ & 0.403 & 0.387 \\
        Carbon stored after storm & $\gamma_{storm}$ & 0.525 & 0.508 \\
        Carbon stored after harvest & $\beta$ & 0.319 & 0.303 \\
        Stem volume parameters:& $v_1$ & 0.0632 & 0.235 \\
        &$v_2$ & -0.0153 & -0.0153 \\
        &$v_3$ & 0.00414 & 0.00621 \\
        &$v_4$ & -0.104 & -0.109 \\
        &$v_5$  & -483 & -1270 
    \end{tabular}
    \end{center}

    \label{parameters}
\end{table}

\subsection{Optimal rotation length}

\begin{figure}[htbp!]
  \centering
    \includegraphics[width=\textwidth]{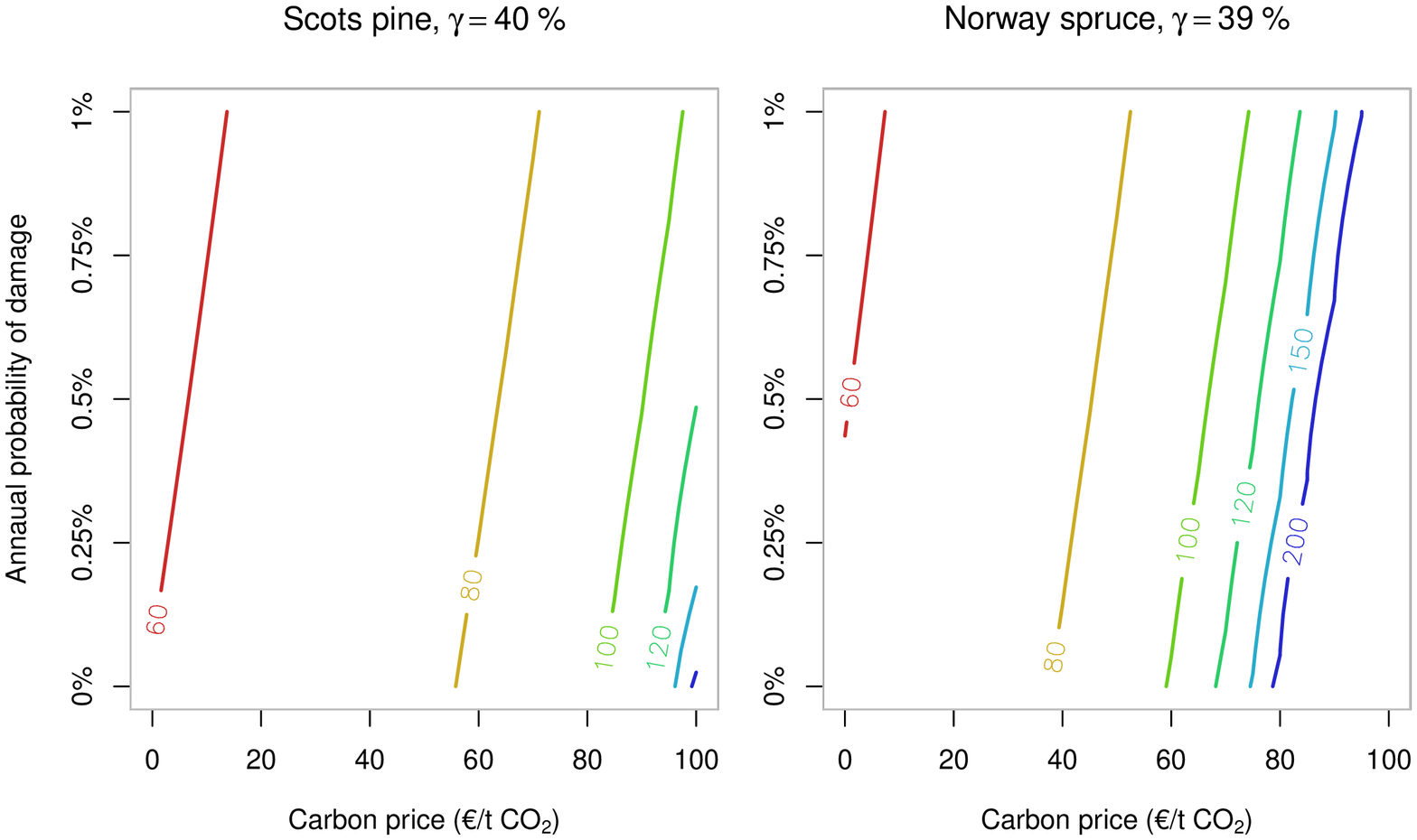}
    \includegraphics[width=\textwidth]{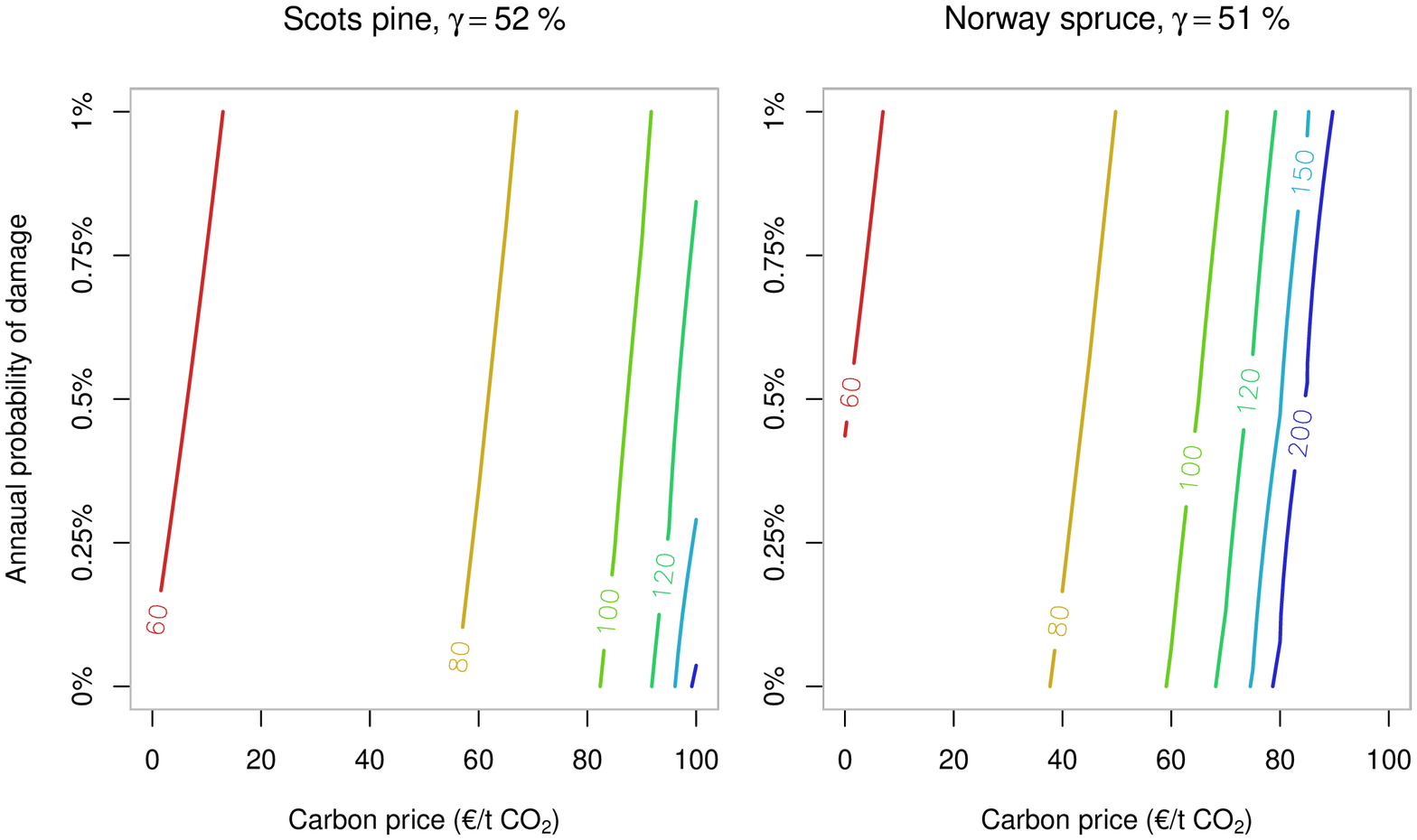}
      \caption{Optimal rotation length (coloured lines, with the labels stating the length in years) for different values of carbon pricing (x-axis) and annual probability of forest damage (y-axis). The top row considers forest fires and the bottom row storm damages.}
      \label{OptT}
\end{figure}

Optimal rotation lengths are presented in Figure \ref{OptT} for Scots pine and Norway spruce under the considered range of carbon prices and damage probabilities, separately for fire and storm damage. Carbon pricing lengthens the optimal rotation considerably. The effect is more pronounced for spruce, which retains its capacity for draining atmospheric carbon longer than pine. 
With high carbon price and low or moderate damage probability, the optimal rotation length can extend to infinity. In such case, despite the declining annual growth rate, the expected net present value from forthcoming carbon sequestration remains higher than the sum of values from bare land and sold timber minus the discounted value of lost carbon stocks. 

A higher damage probability shortens the optimal rotation length, but its impact is far more limited compared to the carbon price. 
Even with the highest considered levels of forest damage probability, the optimal rotations lengthen notably with higher carbon price. 

The correspondence between damage probability and carbon price towards the optimal rotation length is relatively linear with the parameters used in Figure \ref{OptT}. 
The storm and fire cases differ only slightly, the slope of the curves with equal optimal rotation length being more vertical with storms; and henceforth I present the results only for forest fires.
Roughly speaking, an increase of one percentage point in the damage probability corresponds to a decrease of 10 or 15~€/t$_{CO_2}$ in the carbon price, respectively for storm and fire risks; to arrive at the same rotation length.

Under high carbon prices, the optimal rotation length is quite sensitive to how carbon is stored after a harvest or damage, embodied in the parameters $\beta$ and $\gamma$. 
The parameter $\gamma$ affects the situation more -- rather obviously -- if the damage probability is high.
Should more of the carbon be sequestered into long-term storage, thus implying a higher $\beta$, optimal rotation lengths would be shortened considerably. In particular, the infinite rotations would not take place, for it would become preferable to place the harvested carbon from successive rotations in permanent storage, rather than to store it in the growing forest stock.

\subsection{Expected value and risk}

\begin{figure}[htb!]
  \centering
    \includegraphics[width=\textwidth]{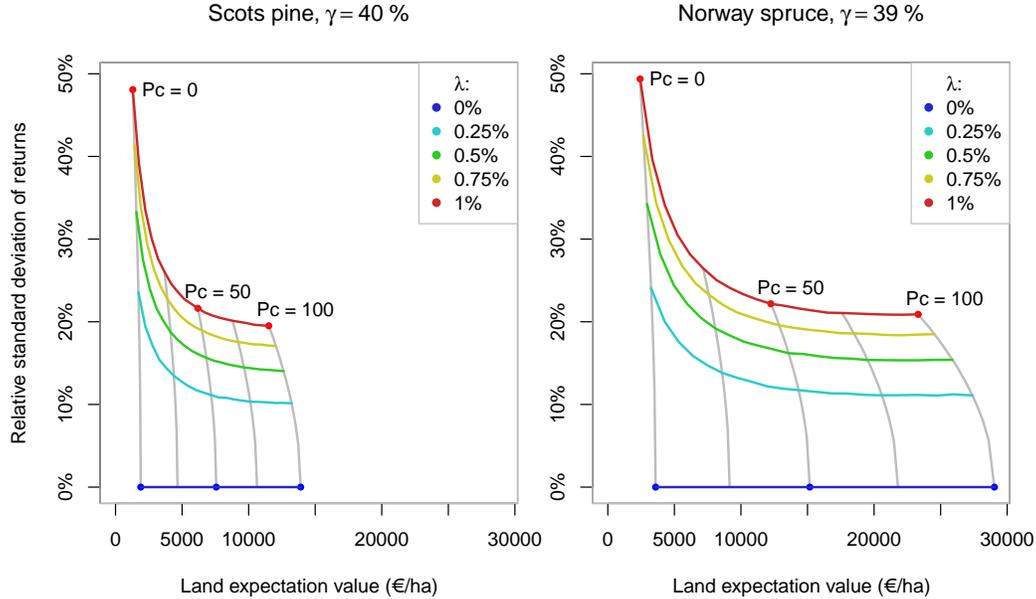}
      \caption{Land expectation value and relative standard deviation of returns for the optimal rotation under different carbon prices (in €/t$_{CO_2}$) and annual damage probabilities $\lambda$ (indicated with colour). The grey lines denote 25~€/$t_{CO_2}$ increments in the carbon price.}
      \label{LEVvsSigma}
\end{figure}

Carbon pricing and damage risks have also a notable impact on the economic returns and risk for the forest owner. Figure \ref{LEVvsSigma} presents the land expectation value (LEV) using equation \eqref{eq:valuefunction} and the relative standard deviation of returns using Monte Carlo sampling.

The land expectation value is 7 to 9 times higher with a carbon price of 100~€/$t_{CO_2}$ than with 0~€/$t_{CO_2}$. Again, this effect is far more pronounced with spruce, which has a higher potential for sequestering carbon. Given this large discrepancy in LEV between the species, and although the two species prefer somewhat differing habitats, carbon pricing could affect species choice while re-planting the forest. The damage risk reduces the LEV somewhat. 

The financial risk is obviously greater with a higher probability for damages.  Carbon pricing, however, effectively mitigates this risk.  For the highest damage probability considered here, the relative risk in returns is effectively halved across the range of carbon prices.  If revenues are solely form harvests, a damage event postpones the final fellings, reducing the returns considerably.

On the other hand, under carbon pricing the forest owner receives revenues early-on from the increasing forest carbon stock, prior to a possible damage event. As the carbon stock is not released immediately to the atmosphere after forest damage, the forest owner is required to pay only a fraction of the forest carbon stock's value in net present value terms. Consequently, the damage risk affects the returns from carbon payments less than from timber sales.  As the bulk of net returns comes from carbon payments under a high carbon price, the relative risk in returns is low.

\subsection{Long-term average carbon stocks}

Whether landowners put their forests towards producing raw material storing carbon depends on the relative pricing between timber and carbon.
To illustrate this more concretely, I calculate the carbon stock and annualized harvests, averaged over a long time horizon in the manner of \citet{Pingoud2018}, for the optimal rotations under different carbon prices and damage probabilities. This can be interpreted as the long-term production possibility frontier between supplying timber and storing carbon, when the rotation length is determined in an economically optimal way.\footnote{Note that the optimal rotation length does not maximize the long-term output due to discounting.}

The average harvest yield with exponentially distributed damages equals the average yield from a single rotation divided with the average length of a single rotation:
\begin{equation}
\label{eq:AvgH}
\frac{e^{-\lambda T^*} v(T^*) }{ \lambda^{-1}(1 - e^{-\lambda T^*} (1+\lambda T^*)) + e^{-\lambda T^*} T^* }.
\end{equation}
The calculation of carbon stocks is based on a Monte Carlo sampling of forest damages, and covers the living tree biomass, dead biomass due to forest damages, and wood products and harvest residues due to final fellings. The carbon stocks for dead biomass and wood products decrease gradually, according to the dynamics described in section \ref{sec:setup}.

\begin{figure}[htb!]
  \centering
    \includegraphics[width=\textwidth]{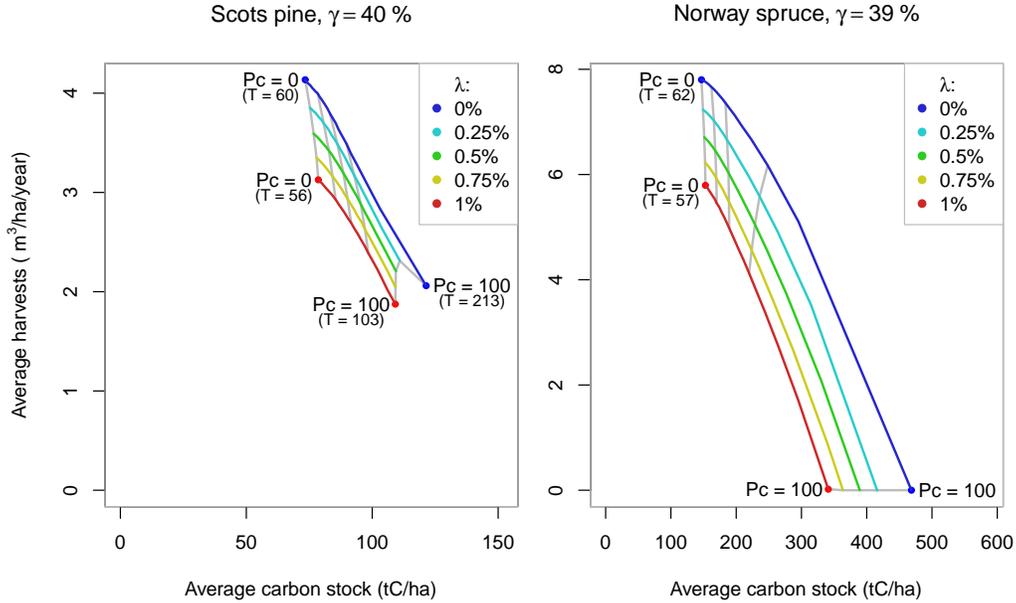}
      \caption{The long-term average of carbon storage (x-axes), including living tree biomass, dead biomass due to damages, wood products and harvest residues; and harvests (y-axis) under different carbon prices (in €/t$_{CO_2}$) and annual damage probabilities $\lambda$ (indicated with colour). The grey lines denote 25~€/$t_{CO_2}$ increments in the carbon price.}
      \label{CvsH}
\end{figure}

The results from these calculations are presented in Figure \ref{CvsH} for the optimal rotation lengths of Figure \ref{OptT}. 
The longer rotations associated with higher carbon prices imply significantly higher carbon stocks and lower harvests, particularly with prices over 75~€/$t_{CO_2}$.
With spruce, the carbon stocks are increased two to three times larger across the range of considered carbon prices, even with the highest probability of forest damages. 

As carbon price increases, there is first a gradual rise in rotation length, followed by a cut-off price after which the optimal rotation length becomes infinite, as was shown in Figure \ref{OptT}.
Concomitantly, harvests can decline rapidly to zero with relatively small changes in the carbon price in Figure \ref{CvsH}.
From the perspective of raw-material production, harvesting decisions seem very sensitive to changes in carbon pricing with higher price levels. One could expect, however, that wood markets would react to such changes by parallel changes in timber prices.

\section{Discussion}

A policy that imposes a price for the flows of forest carbon would incentivize significant climate action in forest management, even in the presence of forest damage risks \citep{Stainback2004,Daigneault2010,Couture2011}. Departing from the earlier studies, I also considered here the fact that the whole carbon stock is not lost immediately after forest damage, even in the case of a forest fire. This lessens the damage risk’s impact on shortening the rotation under carbon pricing, particularly if carbon prices are high.

An efficient policy would require the landowner to effectively pay for the carbon released due to harvests or forest damages. The numerical results reflected this principle, similarly to \citet{Stainback2004} and departing from \citet{Daigneault2010} and \citet{Couture2011}; although the theoretical model allows also alternative specifications.
Regarding forest damages, such policy would provide incentives for reducing the damage risk. 
For harvests, this would incentivize the uses of wood which store the carbon long-term. The practical implementation of who actually pays for the carbon release was left open here, but it was rather assumed that any subsequent release of forest carbon would be factored in the price of timber. Hence, the price $P_F$ was assumed to exclude the value of embodied carbon.

The pricing of forest carbon also increases considerably the revenues from forestry, leading to an increase in the price of land suitable for forest growth. This impact is a clear reflection of the gravity of the climatic problem.
The incentive for afforestation would on one hand effect some of the mitigation potential \citep{Sohngen2003}, but also increase competition for land and lead to spillovers to other land-using sectors, such as agriculture \citep{Golub2009, Doelman2018}.

The result also showed that carbon pricing decreases the relative risk in returns for the landowner, as carbon payments from forest growth generate early-on income during the rotation. With sufficiently high prices for carbon, the optimal rotation can extend to infinity, whereby all revenues for the landowner would arise from draining and storing atmospheric carbon. As a result, timber supply can become highly sensitive on carbon pricing, unless timber prices follow the changes in carbon price.

The analysis presented here did not touch on the discount rate. In financial models, one commonly assumes that riskier investments require higher expected returns, which can be factored in to the net present value calculation through a higher discount rate. The results presented here showed that carbon pricing would reduce the economic risk from forest damages. While this would imply a lower discount rate in itself, one should also consider future volatilities in timber and carbon prices \citep{Chladna2007, Guthrie2009} in order to form a comprehensive picture on the associated risks. 

Another challenge is the construction of meaningful price forecasts multiple decades into the future. For timber prices, some guidance can be achieved through equilibrium modelling of the timber markets \citep[e.g.][]{Sohngen2003}.
This approach allows also to consider how timber markets -- as a whole -- react to carbon prices or other climate policies imposed to forestry.
Carbon prices are prescribed by climate policy; but in an idealized sense depend ultimately on our understanding of the climate problem’s severity, possibly being very volatile \citep{Ekholm2018}. Optimal strategies for climate change mitigation involve increasing carbon prices, which also affects forest management by lengthening the rotation \citep{Ekholm2016}. 

The optimal balance between the supply of carbon storage and renewable raw-material is not hence clear-cut. The answer lies on the relative pricing between carbon and timber.
While the presented results implied a major increase in forest carbon stocks due to a moderate price on carbon -- in itself implying strong contribution to climate action -- the employed static setting did not account for possible price reactions from the timber market. Such responses would return the balance back towards timber production.

Similarly, the extent to which carbon pricing could induce afforestation rests on how fiercely other sectors respond to the increased competition for land \citep{Golub2009, Doelman2018}.
To explore these aspects, future research should further apply dynamic, integrated analyses that go beyond the static-price, single-stand setting used here \citep{Sohngen2003, Favero2017, Siljander2018}. Nevertheless, without a carbon-pricing policy for forests, a major opportunity for mitigating climate change might not be realized.

\section*{Acknowledgements}

The author would like to thank Karoliina Pilli-Sihvola for discussions regarding the problem definition, and Anna Repo, Kim Pingoud and Jari Liski for their help regarding the carbon dynamics. The research has been carried out with funding from the Academy of Finland (decision number 311010).

%
%




\bibliographystyle{elsarticle-harv}
\bibliography{references}

\begin{thebibliography}{28}
\expandafter\ifx\csname natexlab\endcsname\relax\def\natexlab#1{#1}\fi
\expandafter\ifx\csname url\endcsname\relax
  \def\url#1{\texttt{#1}}\fi
\expandafter\ifx\csname urlprefix\endcsname\relax\def\urlprefix{URL }\fi

\bibitem[{Amacher et~al.(2005)Amacher, Malik, and Haight}]{Amacher2005}
Amacher, G.~S., Malik, A.~S., Haight, R.~G., 2005. Not getting burned: the
  importance of fire prevention in forest management. Land Economics 81~(2),
  284--302.

\bibitem[{Chladn{\'{a}}(2007)}]{Chladna2007}
Chladn{\'{a}}, Z., 2007. {Determination of optimal rotation period under
  stochastic wood and carbon prices}. Forest Policy and Economics 9~(8),
  1031--1045.

\bibitem[{Couture and Reynaud(2011)}]{Couture2011}
Couture, S., Reynaud, A., 2011. Forest management under fire risk when forest
  carbon sequestration has value. Ecological Economics 70~(11), 2002--2011.

\bibitem[{Daigneault et~al.(2010)Daigneault, Miranda, and
  Sohngen}]{Daigneault2010}
Daigneault, A.~J., Miranda, M.~J., Sohngen, B., 2010. Optimal forest management
  with carbon sequestration credits and endogenous fire risk. Land Economics
  86, 155--172.

\bibitem[{Doelman et~al.(2018)Doelman, Stehfest, Tabeau, van Meijl, Lassaletta,
  Gernaat, Neumann-Hermans, Harmsen, Daioglou, Biemans, van~der Sluis, and van
  Vuuren}]{Doelman2018}
Doelman, J.~C., Stehfest, E., Tabeau, A., van Meijl, H., Lassaletta, L.,
  Gernaat, D.~E., Neumann-Hermans, K., Harmsen, M., Daioglou, V., Biemans, H.,
  van~der Sluis, S., van Vuuren, D.~P., 2018. {Exploring SSP land-use dynamics
  using the IMAGE model: Regional and gridded scenarios of land-use change and
  land-based climate change mitigation}. Global Environmental Change 48,
  119--135.

\bibitem[{Ekholm(2016)}]{Ekholm2016}
Ekholm, T., 2016. Optimal forest rotation age under efficient climate change
  mitigation. Forest Policy and Economics 62, 62--68.

\bibitem[{Ekholm(2018)}]{Ekholm2018}
Ekholm, T., 2018. Climatic cost-benefit analysis under uncertainty and learning
  on climate sensitivity and damages. Ecological Economics 154, 99 -- 106.

\bibitem[{Favero et~al.(2017)Favero, Mendelsohn, and Sohngen}]{Favero2017}
Favero, A., Mendelsohn, R., Sohngen, B., 2017. {Using forests for climate
  mitigation: sequester carbon or produce woody biomass?} Climatic Change
  144~(2), 195--206.

\bibitem[{Golub et~al.(2009)Golub, Hertel, Lee, Rose, and Sohngen}]{Golub2009}
Golub, A., Hertel, T., Lee, H.~L., Rose, S., Sohngen, B., 2009. {The
  opportunity cost of land use and the global potential for greenhouse gas
  mitigation in agriculture and forestry}. Resource and Energy Economics
  31~(4), 299--319.

\bibitem[{Griscom et~al.(2017)Griscom, Adams, Ellis, Houghton, Lomax, Miteva,
  Schlesinger, Shoch, Siikam{\"{a}}ki, Smith, Woodbury, Zganjar, Blackman,
  Campari, Conant, Delgado, Elias, Gopalakrishna, Hamsik, Herrero, Kiesecker,
  Landis, Laestadius, Leavitt, Minnemeyer, Polasky, Potapov, Putz, Sanderman,
  Silvius, Wollenberg, and Fargione}]{Griscom2017}
Griscom, B.~W., Adams, J., Ellis, P.~W., Houghton, R.~A., Lomax, G., Miteva,
  D.~A., Schlesinger, W.~H., Shoch, D., Siikam{\"{a}}ki, J.~V., Smith, P.,
  Woodbury, P., Zganjar, C., Blackman, A., Campari, J., Conant, R.~T., Delgado,
  C., Elias, P., Gopalakrishna, T., Hamsik, M.~R., Herrero, M., Kiesecker, J.,
  Landis, E., Laestadius, L., Leavitt, S.~M., Minnemeyer, S., Polasky, S.,
  Potapov, P., Putz, F.~E., Sanderman, J., Silvius, M., Wollenberg, E.,
  Fargione, J., 2017. {Natural climate solutions}. Proceedings of the National
  Academy of Sciences 114~(44), 11645--11650.

\bibitem[{Guthrie and Kumareswaran(2009)}]{Guthrie2009}
Guthrie, G., Kumareswaran, D., 2009. Carbon subsidies, taxes and optimal forest
  management. Environmental and Resource Economics 43~(2), 275--293.

\bibitem[{Lehtonen et~al.(2004)Lehtonen, M{\"a}kipää, Heikkinen,
  Siev{\"a}nen, and Liski}]{Lehtonen2004}
Lehtonen, A., M{\"a}kipää, R., Heikkinen, J., Siev{\"a}nen, R., Liski, J.,
  2004. Biomass expansion factors ({BEFs}) for scots pine, norway spruce and
  birch according to stand age for boreal forests. Forest Ecology and
  Management 188~(1), 211 -- 224.

\bibitem[{Lehtonen et~al.(2016)Lehtonen, Ven{\"{a}}la{\"{i}}nen,
  K{\"{a}}m{\"{a}}ra{\"{i}}nen, Peltola, and Gregow}]{Lehtonen2016}
Lehtonen, I., Ven{\"{a}}la{\"{i}}nen, A., K{\"{a}}m{\"{a}}ra{\"{i}}nen, M.,
  Peltola, H., Gregow, H., 2016. {Risk of large-scale fires in boreal forests
  of Finland under changing climate}. Natural Hazards and Earth System Sciences
  16, 239--253.

\bibitem[{Liski et~al.(1998)Liski, Ilvesniemi, M{\"a}kelä, and
  Starr}]{Liski1998}
Liski, J., Ilvesniemi, H., M{\"a}kelä, A., Starr, M., 1998. Model analysis of
  the effects of soil age, fires and harvesting on the carbon storage of boreal
  forest soils. European Journal of Soil Science 49~(3), 407--416.

\bibitem[{Pingoud et~al.(2018)Pingoud, Ekholm, Siev{\"a}nen, Huuskonen, and
  Hynynen}]{Pingoud2018}
Pingoud, K., Ekholm, T., Siev{\"a}nen, R., Huuskonen, S., Hynynen, J., 2018.
  Trade-offs between forest carbon stocks and harvests in a steady state – a
  multi-criteria analysis. Journal of Environmental Management 210, 96--103.

\bibitem[{Pingoud et~al.(2001)Pingoud, Per{\"a}l{\"a}, and
  Pussinen}]{Pingoud2001}
Pingoud, K., Per{\"a}l{\"a}, A.-L., Pussinen, A., 2001. Carbon dynamics in wood
  products. Mitigation and Adaptation Strategies for Global Change 6~(2),
  91--111.

\bibitem[{Pohjola and Valsta(2007)}]{Pohjola2007}
Pohjola, J., Valsta, L., 2007. Carbon credits and management of scots pine and
  norway spruce stands in finland. Forest Policy and Economics 9, 789–--798.

\bibitem[{Reed(1984)}]{Reed1984}
Reed, W.~J., 1984. The effects of the risk of fire on the optimal rotation of a
  forest. Journal of Environmental Economics and Management 11~(2), 180--190.

\bibitem[{Repo et~al.(2011)Repo, Tuomi, and Liski}]{Repo2011}
Repo, A., Tuomi, M., Liski, J., 2011. Indirect carbon dioxide emissions from
  producing bioenergy from forest harvest residues. GCB Bioenergy 3~(2),
  107--115.

\bibitem[{Rogelj et~al.(2018)Rogelj, Popp, Calvin, Luderer, Emmerling, Gernaat,
  Fujimori, Strefler, Hasegawa, Marangoni, Krey, Kriegler, Riahi, {Van Vuuren},
  Doelman, Drouet, Edmonds, Fricko, Harmsen, Havl{\'{i}}k, Humpen{\"{o}}der,
  Stehfest, and Tavoni}]{Rogelj2018}
Rogelj, J., Popp, A., Calvin, K.~V., Luderer, G., Emmerling, J., Gernaat, D.,
  Fujimori, S., Strefler, J., Hasegawa, T., Marangoni, G., Krey, V., Kriegler,
  E., Riahi, K., {Van Vuuren}, D.~P., Doelman, J., Drouet, L., Edmonds, J.,
  Fricko, O., Harmsen, M., Havl{\'{i}}k, P., Humpen{\"{o}}der, F., Stehfest,
  E., Tavoni, M., 2018. {Scenarios towards limiting global mean temperature
  increase below 1.5 °c}. Nature Climate Change 8~(4), 325--332.

\bibitem[{Sedjo and Sohngen(2012)}]{Sedjo2012}
Sedjo, R., Sohngen, B., 2012. {Carbon Sequestration in Forests and Soils}.
  Annual Review of Resource Economics 4~(1), 127--144.

\bibitem[{Seidl et~al.(2017)Seidl, Thom, Kautz, Martin-Benito, Peltoniemi,
  Vacchiano, Wild, Ascoli, Petr, Honkaniemi, Lexer, Trotsiuk, Mairota, Svoboda,
  Fabrika, Nagel, and Reyer}]{Seidl2017}
Seidl, R., Thom, D., Kautz, M., Martin-Benito, D., Peltoniemi, M., Vacchiano,
  G., Wild, J., Ascoli, D., Petr, M., Honkaniemi, J., Lexer, M.~J., Trotsiuk,
  V., Mairota, P., Svoboda, M., Fabrika, M., Nagel, T.~A., Reyer, C. P.~O.,
  2017. {Forest disturbances under climate change.} Nature climate change
  7~(June), 395--402.

\bibitem[{Siljander and Ekholm(2018)}]{Siljander2018}
Siljander, R., Ekholm, T., 2018. {Integrated scenario modelling of energy,
  greenhouse gas emissions and forestry}. Mitigation and Adaptation Strategies
  for Global Change 23~(5), 783--802.

\bibitem[{Sohngen and Mendelsohn(2003)}]{Sohngen2003}
Sohngen, B., Mendelsohn, R., 2003. An optimal control model of forest carbon
  sequestration. American Journal of Agricultural Economics 85~(2), 448--457.

\bibitem[{Stainback and Alavalapati(2004)}]{Stainback2004}
Stainback, G.~A., Alavalapati, J.~R., 2004. Modeling catastrophic risk in
  economic analysis of forest carbon sequestration. Natural Resource Modeling
  17~(3), 299--317.

\bibitem[{Stollery(2005)}]{Stollery2005}
Stollery, K.~R., 2005. Climate change and optimal rotation in a flammable
  forest. Natural Resource Modeling 18~(1), 91--112.

\bibitem[{van Kooten et~al.(1995)van Kooten, Binkley, and
  Delcourt}]{vanKooten1995}
van Kooten, G.~C., Binkley, C.~S., Delcourt, G., 1995. Effect of carbon taxes
  and subsidies on optimal forest rotation age and supply of carbon services.
  American Journal of Agricultural Economics 77~(2), 365--374.

\bibitem[{Xu et~al.(2016)Xu, Amacher, and Sullivan}]{Xu2016}
Xu, Y., Amacher, G.~S., Sullivan, J., 2016. Optimal forest management with
  sequential disturbances. Journal of Forest Economics 24, 106 -- 122.

\end{thebibliography}



\end{document}